\documentclass[preprint2]{aastex}

\def\ltsima{$\; \buildrel < \over \sim \;$}
\def\gtsima{$\; \buildrel > \over \sim \;$}
\def\lsim{\lower.5ex\hbox{\ltsima}}
\def\gsim{\lower.5ex\hbox{\gtsima}}
\def\lapp{\ifmmode\stackrel{<}{_{\sim}}\else$\stackrel{<}{_{\sim}}$\fi}
\def\gapp{\ifmmode\stackrel{>}{_{\sim}}\else$\stackrel{<}{_{\sim}}$\fi}

\setbox0=\hbox{\rm0}
 
\shorttitle{Terzan~5: a third, metal-poor component}
\shortauthors{Origlia et al.}
 
\begin{document} 

\title{The Terzan~5 puzzle: discovery of a third, metal-poor component
\footnote{Based on observations collected at the W.M.Keck
    Observatory, which is operated as a scientific partnership among
    the California Institute of Technology, the University of
    California, and the National Aeronautics and Space Administration.
    The Observatory was made possible by the generous financial
    support of the W. M. Keck Foundation.}}

\author{
L. Origlia\altaffilmark{2},
D. Massari\altaffilmark{3},
R. M. Rich\altaffilmark{4},
A. Mucciarelli\altaffilmark{3},
F. R. Ferraro\altaffilmark{3},
E. Dalessandro\altaffilmark{3},
B. Lanzoni\altaffilmark{3}
}
\affil{\altaffilmark{2} INAF-Osservatorio Astronomico di Bologna,
Via Ranzani 1, I-40127 Bologna, Italy, livia.origlia@oabo.inaf.it}

\affil{\altaffilmark{3} Dipartimento di Fisica e Astronomia, Universit\`a degli Studi
di Bologna, viale Berti Pichat 6/2, 40127 Bologna, Italy}

\affil{\altaffilmark{4} Physics and Astronomy Bldg, 430 Portola Plaza
  Box 951547 Department of Physics and Astronomy, University of
  California at Los Angeles, Los Angeles, CA 90095-1547}


\begin{abstract}
We report on the discovery of 3 metal-poor giant stars in
Terzan 5, a complex stellar system in the the Galactic bulge, 
known to have two populations at [Fe/H]=--0.25 and +0.3.  
For these 3 stars we present new echelle spectra obtained with NIRSPEC at Keck II, 
which confirm their radial velocity membership and provide
average [Fe/H]=--0.79 dex iron abundance and [alpha/Fe]=+0.36 dex enhancement. 
This new population extends the metallicity range of Terzan~5 0.5 dex more metal poor, 
and it has properties consistent with having formed from a gas polluted by core collapse supernovae.
\end{abstract}

\keywords{Galaxy: bulge --- Galaxy: abundances --- stars: abundances
  --- stars: late-type --- techniques: spectroscopic --- infrared:
  stars}

\section{Introduction}
\label{intro}
Terzan 5 is a complex stellar system in the Galactic Bulge. 
It suffers from huge extinction, its average color excess
being E$(B-V)=2.38$ \citep{bar98,val07}, and it is also affected by
significant
differential reddening \citep[$\rm \Delta$E$(B-V)\simeq 0.7$ mag,][]{mas13}.  For
years it has been classified as a globular cluster, although soon
after its discovery its true nature was already disputed \citep[see,
  e.g.,][]{kin72}.

Recent high resolution imaging in the near infrared (IR) obtained with
the Multi-conjugate Adaptive Optics Demonstrator (MAD) at ESO-VLT,
revealed the presence of two distinct red clumps that cannot be
explained by differential reddening or distance effects \citep{fer09}.
Prompt near infrared spectroscopy with NIRSPEC at Keck II demonstrated
that the two stellar populations are characterized by very different
iron abundances ([Fe/H]$=-0.2$ and +0.3).  Subsequent spectroscopic
studies of 33 red giant stars \citep{ori11} fully confirmed the large
metallicity difference between the two populations.  
The sub-Solar component has [Fe/H]$=-0.25\pm0.07$
r.m.s. and it is $\alpha-$enhanced, similar to the
old bulge population that likely formed at early epochs and from a gas 
enriched by a huge amount of type II supernovae (SNe).  The super-Solar 
component, which is possibly a few Gyr younger, has
[Fe/H]$=+0.27\pm0.04$ and approximately Solar [$\alpha$/Fe] abundance ratio,
indicating that it should have originated from a gas polluted by both SNe II and Ia on
a longer timescale.
Both components show a small internal metallicity spread and the most
metal-rich population is also more centrally concentrated \citep{fer09,lan10}.  
These observational facts could be accounted for by 
a proto-Terzan 5 more massive in the past
than today \citep[its current mass being $\rm 2\sim 10^6
M_{\odot}$,][]{lan10}, which possibly experienced at least two
relatively short episodes of star formation with a time delay of a
few Gyr.  

There is an interesting chemical
similarity between Terzan~5 and the bulge stellar population, which
shows a metallicity distribution with two major peaks at sub-Solar and
super-Solar [Fe/H] and a third, minor component with significantly
lower ([Fe/H]$\approx -1.0$) metallicity \citep[see e.g.][ and
references therein]{zoc08,hil11,joh11,ric12,utt12,ness13a,ness13b}.
These bulge stellar populations show [$\alpha$/Fe] enhancement up to
about Solar [Fe/H], and then a progressive decline towards Solar
values at super-Solar [Fe/H]. 
Such a trend is at variance either with the one observed in the
thick disk, where the knee occurs at significantly lower values of
[Fe/H], and with the rather flat distribution of the thin disc with
about Solar [$\alpha$/Fe]. 
Chemical abundances of bulge dwarf stars
from microlensing experiments \citep[see e.g.][and references
therein]{coh10,ben13} also suggest the presence of two populations,
a sub-Solar and old one with [$\alpha$/Fe] enhancement, and a possibly
younger, more metal-rich one with decreasing [$\alpha$/Fe] enhancement
with increasing [Fe/H].
 
This Letter presents the discovery of 3 red giant stars belonging to
Terzan~5, with metallicity far below the sub-Solar component observed so far.

\section{Observations and chemical abundance analysis}
In the context of an ongoing
spectroscopic survey 
with VLT-FLAMES and Keck-DEIMOS 
of the Terzan 5 stellar populations, 
aimed at constructing a massive 
database of radial velocities and metallicities (Massari et al.; Ferraro et al.; 2014 in preparation),
we found some indications of the presence of a minor
($\sim 3\%$) component significantly more metal-poor than the
sub-Solar population of Terzan~5.  
We acquired high
resolution spectra of 3 radial velocity candidate metal-poor giants 
members of Terzan~5.
Observations using NIRSPEC \citep{ml98} at Keck II  
were undertaken on 17 June 2013.   
We used the
NIRSPEC-5 setting to enable observations
in the $H$-band and a
$0.43\arcsec$ slit width that provides an overall spectral resolution
R=25,000. 

\begin{figure*}[!hp]
\begin{center}
\includegraphics[scale=0.7]{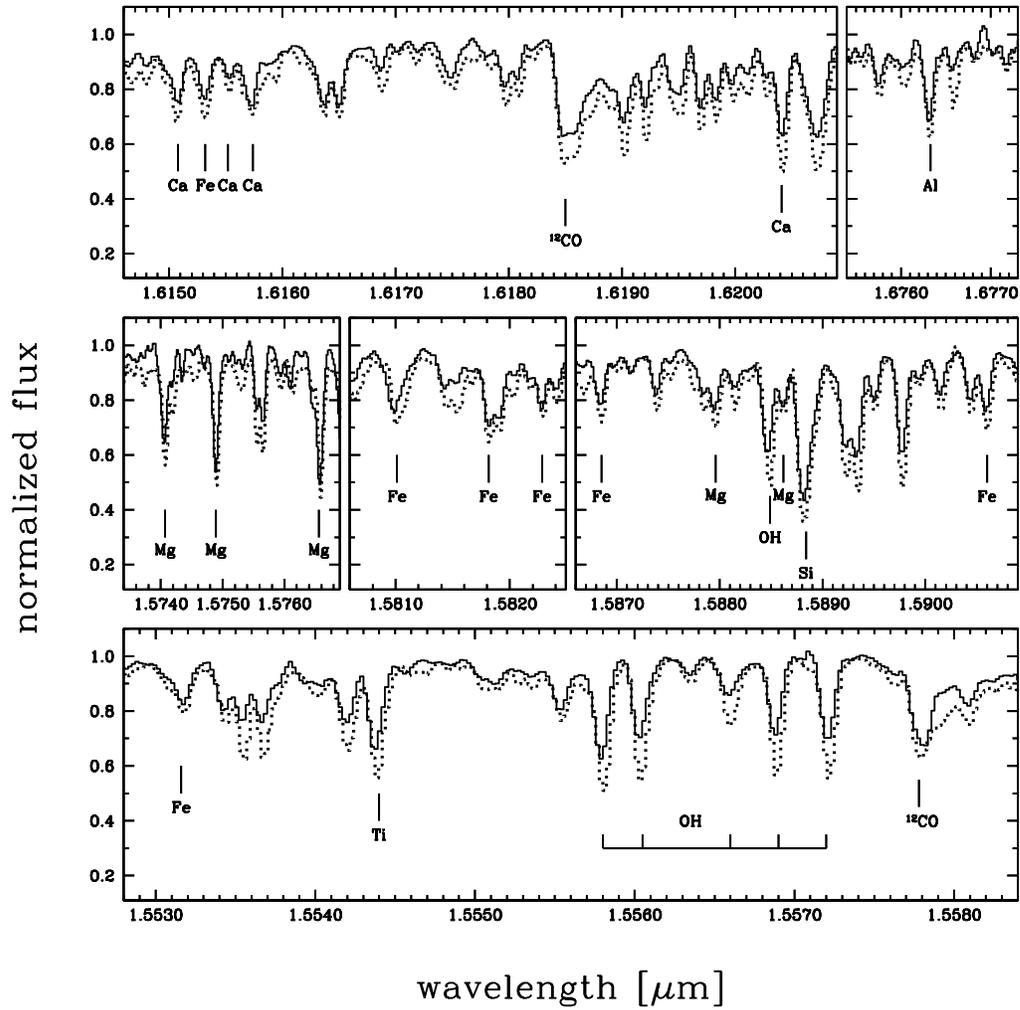}
\caption{Portion of the NIRSPEC $H$-band
  spectra of two red giants of Terzan~5 with similar temperature ($\rm
  T_{eff}\approx 3800$ K), but different chemical abundance patterns
  (solid line for the metal-poor star \#243, dotted line for a
  sub-Solar star at [Fe/H]$\approx$-0.22 from \citealp{ori11}).  
  The metal poor giant \#243 has significantly shallower features. 
  A few atomic lines and molecular bands of interest are marked.}
\label{spec}
\end{center}
\end{figure*}

Data reduction has been performed by using 
the REDSPEC IDL-based package developed at the UCLA IR Laboratory.
Each spectrum has been sky subtracted by using nod pairs, corrected for flat-field  
and calibrated in wavelength using arc lamps.
An  O-star spectrum observed during the same night has been used to remove
to check and remove telluric features. 
The signal to noise ratio iper rersolution element of the final spectra is always $>$30. 
Figure~\ref{spec} shows portions of the observed spectra and the comparison with 
a Terzan~5 giant with similar stellar parameters and higher iron content from the sub-Solar population 
studied by \citet{ori11}.

We compare the observed spectra with synthetic ones 
and we obtain accurate chemical abundances of C and O using molecular lines and 
of Fe, Ca, Si, Mg, Ti and Al using neutral atomic lines, as also 
described in \citet{ori11} and references therein.

We made use of both spectral synthesis analysis and equivalent width measurements 
of isolated lines.
Synthetic spectra covering a wide range of stellar parameters and elemental abundances  
have been computed by using 
the same code as in \citet{ori11} and described in detail
in \citet{ori02} and \citet{ori04}. 
The code uses the LTE approximation, the 
molecular blanketed model atmospheres of
\citet{jbk80} at temperatures $\le $4000~K, and 
the \citet{gv98} abundances for the Solar reference.

Stellar temperatures have been first estimated from colors, by using
the reddening estimates by \citet{mas13} and the color-temperature
scale by \citet{mon98}, calibrated on globular
cluster giants.  Gravity has been estimated from theoretical
isochrones \citep{pie04,pie06}, according to the position of the stars on the red
giant branch (RGB). An average microturbulence velocity of 2 km/s has been adopted 
\citep[see e.g.][ for a detailed discussion]{ori97}.  
The simultaneous spectral fitting of the CO and OH molecular
lines 
that are especially sensitive to temperature, gravity and
microturbulence variations \citep[see also][]{ori02},
allow us to fine-tune our best-fit adopted stellar parameters.

\section{Results}
Our provisional estimate
for the systemic velocity of Terzan~5, as inferred from
our VLT-FLAMES and Keck-DEIMOS survey,  
is --82 km/s  with a velocity dispersion of $\approx$15~km/s.

From the NIRSPEC spectra we first measured 
the radial velocity of the 3 stars under study and confirm values within $\approx 1\sigma$ 
from the systemic velocity of Terzan~5 (see Table~\ref{tab1}).
These stars are located in the central region of Terzan~5, at distances between 13 and 71 arcsec 
from the center (see Table~\ref{tab1}). 
Our VLT-FLAMES and Keck-DEIMOS survey shows that in 
this central region the contamination by field stars with similar radial velocities 
and metallicity is negligible (well below 1\%).
Preliminary analysis of proper motions also indicates that these stars are likely members of Terzan~5.

We then measured the chemical abundances of iron, 
alpha-elements, carbon and aluminum.  
Our best-fit estimates of the stellar temperature and gravity,  
radial velocity and chemical abundances with 
$\rm 1\sigma$ random errors are listed in Table~\ref{tab1}.  
In the evaluation of the overall error budget we also estimate
that systematics due to 
$\rm \Delta T_{eff}\pm$200~K, 
$\rm \Delta log~g\pm$0.5 dex, $\rm \Delta \xi\pm$0.5 km/s variations in the adopted stellar parameters can affect 
the inferred abundances by $\approx \pm 0.15$ dex. 
However, the derived abundance ratios 
are less dependent 
on the systematic error, since 
most of the spectral features used to measure abundance ratios have similar trends with varying 
the stellar parameters, and 
at least some degeneracy between abundance and the latter is canceled out.

We find the average iron abundance [Fe/H]=--0.79$\pm$0.04 r.m.s.
to be significantly lower (by a factor of $\sim 3$) than the
value of the sub-Solar population 
([Fe/H]$=-0.25$), pointing towards the presence of 
a distinct population in Terzan~5, rather than
to the low metallicity tail of the sub-Solar component.

\begin{figure*}
\begin{center}
\includegraphics[scale=0.7]{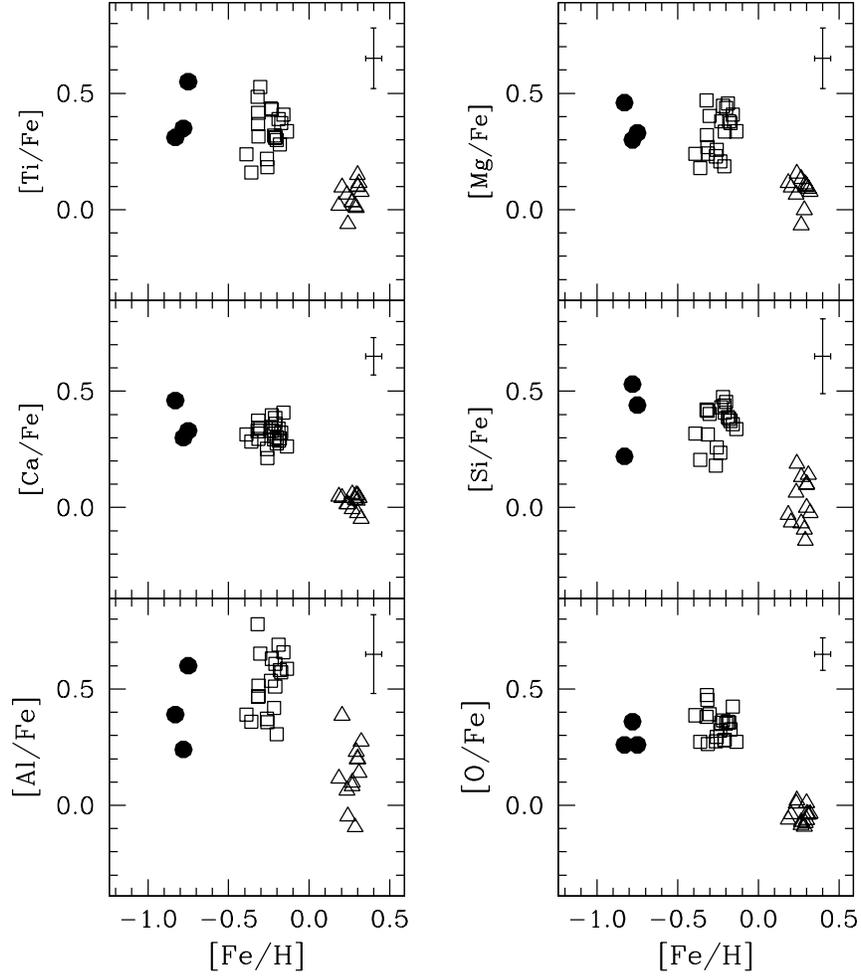}
\caption{Individual [$\alpha$/Fe] and [Al/Fe] abundance ratios as a function of
  [Fe/H] for the 3 observed metal-poor giants (solid dots), and the 20 sub-Solar (open squares) and 13 super-Solar 
  (open triangles) giants from \citet{ori11}, for comparison.
  Typical errorbars are plotted in the top-right corner of each panel.  
  }
\label{alpha}
\end{center}
\end{figure*}

As shown in Figure~\ref{alpha}, our newly discovered metal-poor 
population has an average $\alpha$-enhancement ([$\alpha$/Fe]$=
+0.36 \pm 0.04$ r.m.s.) similar to that of the sub-Solar one, 
indicating that both populations likely formed early
and on short timescales from a gas polluted by type II SNe.  

As the stars belonging to the sub-Solar component,
also these other giants with low iron content show an enhanced
[Al/Fe] abundance ratio (average [Al/Fe]$=+0.41 \pm 0.18$ r.m.s.) and
no evidence of Al-Mg and Al-O anti-correlations, and/or large [O/Fe]
and [Al/Fe] scatters, although no firm conclusion can be drawn with
3 stars only.

We also measured some [C/Fe] depletion 
(at least in stars \#243 and \#262),
as commonly found in giant stars and explained with mixing
processes in the stellar interiors during the evolution along the RGB.

\section{Discussion and Conclusions}

New spectroscopic observations of 3 stars, members of
Terzan~5, have provided a further evidence of the complex nature of
this stellar system and of its likely connection with the bulge
formation and evolution history.

We find that Terzan~5 hosts a third, metal-poorer population with
average [Fe/H]=$-0.79 \pm 0.04$ r.m.s. and [$\alpha$/Fe] enhancement.
From our VLT-FLAMES/Keck-DEIMOS survey,
we estimate that this component represents a minor fraction (a few percent) of 
the stellar populations in Terzan~5. 

Notably, a similar fraction ($\approx 5\%$) of metal-poor stars ([Fe/H]$\approx
-1$) has been also detected in the bulge \citep[see e.g.][and
references therein]{ness13a,ness13b}. 
This metal-poor population shows a kinematics
typical of a slowly rotating spheroidal or a metal weak thick disk
component.  

Our discovery significantly enlarges the metallicity range
covered by Terzan~5, which amounts to $\Delta$[Fe/H]$\approx 1$ dex.
Such a value is completely unexpected
and unobserved in genuine globular clusters. Indeed, within the Galaxy
only another globular-like system, namely $\omega$Centauri, 
harbors stellar populations with a
large ($>$1~dex) spread in iron \citep{norris95, sollima05, johnson10, pancino11}. 
This evidence strongly sets Terzan 5 and
$\omega$Centauri apart from the class of genuine globular clusters,
and suggests a more complex formation and evolutionary history for these two
multi-iron systems. 

It is also interesting to note that detailed
spectroscopic screening recently performed in $\omega$Centauri
revealed an additional sub-component 
significantly more metal-poor (by $\Delta$[Fe/H]$\sim 0.3-0.4$ dex)
than the dominant population \citep{pancino11}. 
The authors suggest that this is best accounted for in 
a self-enrichment scenario, where
these stars could be the remnants of the fist stellar generation in
$\omega$Centauri.  

The three populations of Terzan~5 
may also be explained with some self-enrichment.
The narrow peaks in their metallicity distribution can be the result of  
a quite bursty star formation activity in the proto-Terzan~5, which should
have been much more massive in the past to retain the SN ejecta and
progressively enrich in metals its gas.  However, Terzan~5 might also
be the result of an early merging of fragments with sub-Solar metallicity
at the epoch of the bulge/bar formation, and with younger and more
metal-rich sub-structures following subsequent interactions with the
central disk.

However, apart from the similarity in terms of large iron range and possible 
self-enrichment, $\omega$Centauri and Terzan 5 likely had quite different origins 
and evolution.
It is now commonly accepted that $\omega$Centauri can be the remnant of a dwarf
galaxy accreted from outside the Milky Way \citep[e.g.][]{bekki03}.
At variance, the much higher metallicity of Terzan~5 and its chemical similarity 
to the bulge populations suggests some \emph{symbiotic} evolution
between these two stellar systems.

\acknowledgements This research was supported by the Istituto
Nazionale di Astrofisica (INAF, under contract PRIN-INAF 2010).  
The research is also part of the project {\it COSMIC-LAB}
(http://www.cosmic-lab.eu) funded by the {\it European Research
Council} (under contract ERC-2010-AdG-267675).  The authors wish to
recognize and acknowledge the very significant cultural role and
reverence that the summit of Mauna Kea has always had within the
indigenous Hawaiian community.  We are most fortunate to have the
opportunity to conduct observations from this mountain.
The authors wish to thank the anonymous Referee for his/her useful comments.

\newpage

\begin{deluxetable}{lllclcrlllllllc}
\tabletypesize{\scriptsize} \tablecaption{Stellar parameters and
abundances for the 3 observed giants in Terzan~5.}
\tablewidth{0pt} \tablehead{ 
\colhead{\#} & 
\colhead{RA (2000)}&
\colhead{Dec (2000)}& 
\colhead{$\rm T_{eff}$}& 
\colhead{log~g}&
\colhead{$\rm v_r^a$}&
\colhead{r$^b$}&
\colhead{$\rm [Fe/H]$}& 
\colhead{$\rm [O/Fe]$}& 
\colhead{$\rm [Si/Fe]$}& 
\colhead{$\rm [Mg/Fe]$}&
\colhead{$\rm [Ca/Fe]$}& 
\colhead{$\rm [Ti/Fe]$}& 
\colhead{$\rm [Al/Fe]$}& 
\colhead{$\rm [C/Fe]$} } \startdata
243 &267.0088362 &-24.7951362 & 3800 & 1.0   & -74 &71  & -0.78  &+0.36  &+0.53  &+0.30  &+0.38  &+0.35 &+0.24  &-0.12  \\
    &&&&&&  &$\pm$0.02  &$\pm$0.05  &$\pm$0.10  &$\pm$0.03  &$\pm$0.04  &$\pm$0.10  &$\pm$0.10  &$\pm$0.07\\
262 &267.0210698 &-24.7755223 & 4000 & 1.0   &-64  &13  &-0.83  &+0.26  &+0.22  &+0.46  &+0.39  &+0.31  &+0.39  &-0.47  \\
    &&&&&&  &$\pm$0.08  &$\pm$0.09  &$\pm$0.13  &$\pm$0.08  &$\pm$0.08  &$\pm$0.13  &$\pm$0.13  &$\pm$0.11\\
284 &267.0194897 &-24.7724098 & 3800 & 0.5   &-92 &24    &-0.75  &+0.25  &+0.44  &+0.33  &+0.36  &+0.55  &+0.60  &-0.05  \\
    &&&&&&  &$\pm$0.05  &$\pm$0.08  &$\pm$0.11  &$\pm$0.13  &$\pm$0.08  &$\pm$0.11  &$\pm$0.11  &$\pm$0.09\\
\enddata                       
\tablenotetext{a}{Heliocentric radial velocity in $\rm km~s^{-1}$.}
\tablenotetext{b}{Radial distance from the center of Terzan~5 in arcsec.}
\label{tab1}
\end{deluxetable}

\begin{thebibliography}{} 

\bibitem[Barbuy et al.(1998)]{bar98} 
Barbuy, B., Bica, E., \& Ortolani, S.\ 1998, \aap, 333, 117
\bibitem[Bensby et al.(2013)]{ben13}
Bensby, T. et al. 2013, \aap, 549, 147
\bibitem[Bekki \& Freeman(2003)]{bekki03} Bekki, K., \& Freeman,
  K.~C.\ 2003, \mnras, 346, L11
\bibitem[Cohen et al.(2010)]{coh10}
Cohen, J.G., Gould, A., Thompson, I.B., Feltzing, S., Bensby, T., Johnson, J.A., Huang, W.,
Melendez, J., Lucatello, S., \& Asplund, M. 2010, \apj, 711, L48
\bibitem[Ferraro et al.(2009)]{fer09} 
Ferraro, F.R., et al., 2009, Nature, 462, 483
\bibitem[Grevesse \& Sauval(1998)]{gv98} 
Grevesse, N., \& Sauval, A. J. 1998, {\em Space Science Reviews}, 85, 161
\bibitem[Hill et al. (2011)]{hil11}
Hill, V., Lecureur, A., Gomez, A., Zoccali, M., Schultheis, M., Babusiaux, C., Royer, F.,
Barbuy, B., Arenou, F., Minniti, D., \& Ortolani, S. 2011, \aap, 535, 80
\bibitem[Johnson, Bernat \& Krupp(1980)]{jbk80}
 Johnson, H. R., Bernat, A. P., \& Krupp, B. M. 1980, ApJS, 42, 501
\bibitem[Johnson \& Pilachowski(2010)]{johnson10} Johnson,
  C.~I., \& Pilachowski, C.~A.\ 2010, \apj, 722, 1373
\bibitem[Johnson et al. (2011)]{joh11}
Johnson, C.I., Rich, R.M., Fulbright, J.P, Valenti, E., \& McWilliam, A. 2011, \apj, 732, 108
\bibitem[King(1972)]{kin72}
King, I.R. 1972, \aap, 19, 166
\bibitem[Lanzoni et al.(2010)]{lan10} 
Lanzoni, B., et al., 2010, \apj, 717, 653
\bibitem[McLean(1998)]{ml98} 
McLean, I. et al. 1998, SPIE, 3354, 566
\bibitem[Massari et al.(2013)]{mas13}
Massari, D., Mucciarelli, A., Dalessandro, E., Ferraro, F.R., Origlia, L., Lanzoni, B.,
Beccari, G., Rich, R.M., Valenti, E., \& Ransom, S.M. 2012, \apj, 755, L32
\bibitem[Montegriffo et al.(1998)]{mon98} 
Montegriffo, P., Ferraro, F.R., Fusi Pecci, F., \& Origlia, L., 1998, \mnras, 297, 872
\bibitem[Ness et al.(2013a)]{ness13a}
Ness, M., Freeman, K., Athanassoula, E., Wylie-De-Boer, E., Bland-Hawthorn, J.,
Asplund, M., Lewis, G.F., Yong, D., Lane, R.R., \& Kiss, L.L. 2013a, \mnras, 430, 836
\bibitem[Ness et al.(2013b)]{ness13b}
Ness, M., Freeman, K., Athanassoula, E., Wylie-De-Boer, E., Bland-Hawthorn, J.,
Asplund, M., Lewis, G.F., Yong, D., Lane, R.R., Kiss, L.L., \& Ibata, R. 2013b, \mnras, 432, 2092
\bibitem[Norris \& Da Costa(1995)]{norris95} Norris, J.~E., \& Da
  Costa, G.~S.\ 1995, \apj, 447, 680
\bibitem[Origlia, Moorwood \& Oliva(1993)]{OMO93} 
Origlia, L., Moorwood, A. F. M., \& Oliva, E. 1993, \aap, 280, 536
\bibitem[Origlia et al.(1997)]{ori97} 
Origlia, L., Ferraro, F. R., Fusi Pecci, F., \& Oliva, E. 1997, \aap, 321, 859
\bibitem[Origlia, Rich \& Castro(2002)]{ori02} 
Origlia, L., Rich, R. M., \& Castro, S. 2002, \aj, 123, 1559
\bibitem[Origlia \& Rich(2004)]{ori04} 
Origlia, L., \& Rich, R. M. 2004, \aj, 127, 3422
\bibitem[Origlia et al.(2011)]{ori11} 
Origlia, L., Rich, R.M., Ferraro, F.R., Lanzoni, B., Bellazzini, M.,
Dalessandro, E., Mucciarelli, A., Valenti, E., Beccari, G. 2011, \apj, 726, L20
\bibitem[Pancino et al.(2011)]{pancino11} Pancino, E., Mucciarelli,
  A., Sbordone, L., et al.\ 2011, \aap, 527, A18
\bibitem[Pietrinferni et al.(2004)]{pie04}
Pietrinferni, A., Cassisi, S., Salaris, M., \& Castelli, F. 2004, \apj, 612, 168
\bibitem[Pietrinferni et al.(2006)]{pie06}
Pietrinferni, A., Cassisi, S., Salaris, M., \& Castelli, F. 2006, \apj, 642, 797
\bibitem[Rich, Origlia \& Valenti(2012)]{ric12}
Rich, R.M., Origlia, L., \& Valenti, E. 2012, \apj, 746, 59
\bibitem[Sollima et al.(2005)]{sollima05} Sollima, A., Pancino, E.,
  Ferraro, F.~R., et al.\ 2005, \apj, 634, 332
\bibitem[Uttenthaler et al.(2012)]{utt12} Uttenthaler, S., Schultheis,
  M., Nataf, D.M., Robin, A.C., Lebzelter, T., \& Chen, B. 2012, \aap,
  546, 57
\bibitem[Valenti et al.(2007)]{val07} 
Valenti, E., Ferraro, F. R., \& Origlia, L., \ 2007, \aj, 133, 1287
\bibitem[Zoccali et al.(2008)]{zoc08} 
Zoccali, M., Hill, V., Lecureur, A., Barbuy, B., Renzini, A., Minniti, D., G{\'o}mez, A., \& Ortolani, S.\ 2008, \aap, 486, 177

\end{thebibliography}
\end{document}